\documentclass[twocolumn, amsmath, superscriptaddress, amsfonts,prb]{revtex4}
\pdfoutput=1
\usepackage{graphicx}
\usepackage{lipsum}

\begin{document}

\title{Whisker growth on Sn thin film accelerated under gamma-ray irradiation}
\author{Morgan Killefer}\affiliation{Department of Physics and Astronomy, University of Toledo, Toledo,OH 43606}%\email{Morgan.Killefer@rockets.utoledo.edu}
\author{Vamsi Borra}\affiliation{Department of Electrical Engineering and Computer Science, University of Toledo, Toledo, OH 43606} %\email{vamsi.borra@rockets.utoledo.edu}
\author{Ahmed Al-Bayati}\affiliation{Department of Physics and Astronomy, University of Toledo, Toledo,OH 43606}%\email{Morgan.Killefer@rockets.utoledo.edu}
\author{Daniel G. Georgiev}\affiliation{Department of Electrical Engineering and Computer Science, University of Toledo, Toledo, OH 43606}%\email{daniel.georgievv@utoledo.edu}
\author{Victor G. Karpov}\affiliation{Department of Physics and Astronomy, University of Toledo, Toledo,OH 43606}%\email{victor.karpov.utoledo.edu}
\author{E. Ishmael Parsai}\affiliation{Department of Radiation Oncology, University of Toledo Health Science Campus, Toledo, OH 43614}%\email{ishamel.parsai@utoledo.edu}
\author{Diana Shvydka}\email{diana.shvydka@utoledo.edu}\affiliation{Department of Radiation Oncology, University of Toledo Health Science Campus, Toledo, OH 43614}
\begin{abstract}
We report on the growth of tin (Sn) metal whiskers that is significantly accelerated under gamma-ray irradiation. The studied Sn thin film, evaporated on glass substrate, was subjected to a total of 60 hours of irradiation over the course of 30 days. The irradiated sample demonstrated enhanced whisker development, in both densities and lengths, resulting in an acceleration factor of $\sim$50. We attribute the observed enhancement to gamma-ray induced electrostatic fields, affecting whisker growth kinetics. These fields are due to the substrate charging under ionizing radiation of gamma-rays. We propose that gamma-ray irradiation can be a tool for accelerated testing of whisker propensity.
\end{abstract}

\maketitle

{\it Introduction}. Understanding of mechanisms of metal whiskers growth on technologically important metals such as Sn presents a long-standing challenge. As a consequence, unpredictable catastrophic device failures afflicting multiple industries from aerospace to automotive \cite{NASA,brusse} and costing billions of dollars raise significant reliability concerns. Whisker mitigating techniques, developed over several decades of research remain disputable; moreover the absence of reliable accelerated life testing procedures makes it especially difficult to evaluate their effectiveness with tests limited in time, such as those suggested by Joint Electron Devices Engineering Council.\cite{JEDEC201,NASAcom}

One of the approaches recently developed by our group may offer insights into the whisker growth mechanism, and offer a possibility of accelerated test development: we have found that whisker growth can be significantly accelerated when metal films on insulating substrates are subjected to high-energy (6MeV) electron beams. \cite{vasko1} We have shown that the  acceleration factors (formally defined below) of  $\sim$200 result from static electric fields generated due to accumulation of charged defects in the insulating substrates thus subjecting deposited metal films to the electric fields perpendicular to the film surface. The observed effect was explained by the electrostatic theory of metal whisker growth. \cite{karpov,shvydka}  Use of ionizing radiation for generation of electric fields may be preferable to a simpler capacitor-type setup \cite{vasko2} which requires a second electrode, often leading to shorts due to whisker growth through the capacitor air gap.

In the present work we irradiate thin-film tin samples deposited on glass substrate with an Ir-192 gamma-ray source. The choice of the source rules out any radiation effects within the metal film due to its small thickness. At the same time, atomic ionizations are the only radiation effects within the substrate, where high-energy photons are very unlikely to produce radiation damage via atomic displacements, \cite{oen} but create charged defects through ionization.

{\it Experimental details.} Thin-film Sn samples were deposited on 3-mm thick soda-lime glass substrates coated with transparent conducting oxide (TCO, specifically, SnO$_2$:F with nominal 15 Ohm/square sheet resistance; TEC-15 glass from Pilkington). This type of substrate, routinely used in our labs for thin-film photovoltaic device fabrication, supports good film adhesion with various deposition applications. We followed our general substrate preparation procedure where substrates cut into 3x6 cm$^2$ pieces are washed in cleaning solution (Micro-90), then rinsed with de-ionized (DI) water, followed with ultra-sonication bath in acetone for 20-25 min, and, finally, with ultra-sonication bath in DI water for 20-25 min. In between these steps, the surfaces are blown dry with nitrogen.

Sn samples were prepared by vacuum evaporation of 99.999\% pure Sn (from Kurt J. Lesker) from tungsten boats onto the substrates; the detailed fabrication process steps are described in Ref. \onlinecite{borra}. Sn layers of 250$\pm$25nm thickness were deposited at a growth rate of 2-4 \AA/s through stainless steel masks, resulting in a set of 5 mm wide Sn strips over continuous TCO coating. The deposition was conducted simultaneously for all samples.

\begin{figure}[h]
\includegraphics[width=0.40\textwidth]{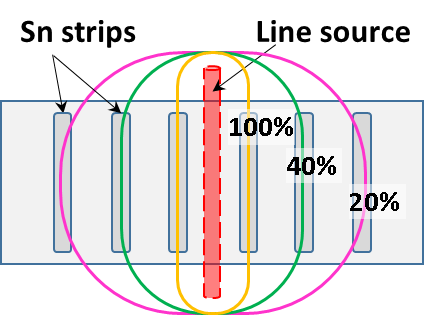}
\caption{A sketch of the sample irradiation geometry, showing the line source position (red dashed cylinder in the middle); a dose fall-off from a line source follows 1/r law, shown with labeled isodose lines at locations of metallized Sn strips.  \label{fig:setup}}.
\end{figure}

For irradiation, we used gamma-ray source of Ir-192 clinical high-dose rate (HDR) afterloader (VariSource, Varian Medical Systems) at UTMC Dana Cancer Center facility. That source is poly-energetic, with the average photon energy $\sim$380keV (beta particles emitted by the source are fully absorbed by Ni-Ti alloy encapsulating the radionuclide core), and $\sim$74 days half-life. \cite{borg} The corresponding photon attenuation in Sn film is negligible, while the effective attenuation coefficient in glass   $\mu\sim$ 0.24cm$^{-1}$, resulting in $\exp (- \mu x)\approx 7$\% attenuation in a substrate over x=0.3 cm thickness. The source of linear dimension of $\sim$3 mm, was used at multiple dwell positions 5 mm apart along a straight line, which makes it effectively a line source positioned through the center of the sample, parallel to its shorter dimension, as shown in Fig. \ref{fig:setup}.

The dose deposition in glass falls off with the distance r from the source as 1/r, \cite{jons} resulting in dose distribution (at the positions of Sn strips) represented by isodose lines in Fig. 1. Samples were irradiated to two dose levels of 10kGy and 20kGy (Gy, "gray", is a derived unit of ionizing radiation dose in the SI system, defined as the absorption of one joule of radiation energy per kilogram of matter 1Gy=1 J/kg) at the position of 100\% isodose line. The corresponding HDR source dwell times were calculated using BrachyVision treatment planning system. The specific choice of the dose levels was made based on the data from our electron beam irradiation experiments, \cite{vasko1} where samples receiving 20kGy showed accelerated whisker growth.

Achieving the above dose levels in the environment of busy medical facility required multiple irradiation sessions, delivered on average over several days (approximately 2-4 hours per 1kGy of dose), with longer delivery times necessary to make up for the source decay from its maximum  activity of 10Ci at the beginning of irradiations. As an example, 10kGy would typically take irradiation time $t_R\sim 30$ hours. The control sample was stored under lab conditions for the duration of experiments.

A scanning electron microscope (SEM), Hitachi S-4800, was utilized as a primary characterization tool for metal film surface imaging. It was operated in secondary-electron mode with acceleration voltage of 5 kV in order to limit the observations to the film surface. SEM imaging of the irradiated sample was conducted by the following schedule: before irradiation (0 kGy), after receiving 10 and 20 kGy total dose, and finally 30 days after the last irradiation (20kGy +30 days shelf). The final round of irradiations and image acquisitions were finished about 30 days after samples deposition, thus the last imaging set (after 30 additional days of shelf) was acquired about 60 days after sample deposition.

For imaging, we identified 3 regions on each side of the line source: central, closest to the source and corresponding to dose levels of 100\%, and two additional regions on each side, corresponding to average dose levels of 40, and 20\% (see Figure 1). For each region we collected 40 SEM images per dose level (0, 10, and 20kGy) and 30 days of shelf life after the final irradiation. Metal whiskers were counted in all collected images, and the ImageJ software package was used to measure the lengths of all metal whiskers identified on sample surfaces. The whisker length was measured as a sum of straight portions for whiskers with complex shapes. The control sample was imaged twice: first, 35 days, and then 60 days after deposition. For the irradiated sample these time intervals corresponded to 1) receiving the total dose of 20kGy, and 2) 20kGy + 30 days of shelf life. We remind, however, that the actual time under gamma-rays delivered in sessions, was about 60 hours to achieve 20kGy.

\begin{figure*}[ht]
\centering
\includegraphics[width=0.70\textwidth]{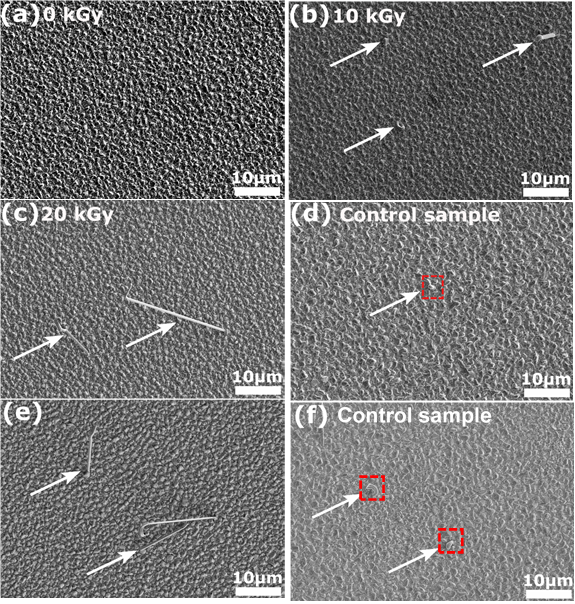}
\caption{SEM images of: (a) sample before irradiation; (b) after exposure to 10kGy dose; here three short whiskers are visible. (c) Longer metallic whiskers grew after 20kGy delivered dose; this picture shows two longer than average representatives. (d) Very small whisker (highlighted in red box) imaged on control sample after 35 days of sample deposition; (e) Three whiskers are shown on the surface of the same sample as in (c) after additional 30 days of shelf life (60 days post-deposition); (f) Control sample 60 days post-deposition with two whiskers visible in the imaged region. \label{fig:images}}.
\end{figure*}

{\it Results.} SEM scans of samples before irradiation, conducted within a week from sample deposition, revealed no whiskers; a representative image is shown in Fig. \ref{fig:images}(a). Whiskering was observed after irradiation to 10kGy [Fig. \ref{fig:images}(b)], further enhanced after the final dose of 20kGy was delivered [Fig. \ref{fig:images}(c), showing a set of two longer than average whiskers]. The first round of irradiation (to 10kGy) was finished roughly 15 days after sample depositions, the second round (to 20kGy) - about 30 days after sample deposition, at this stage we also imaged the control sample, finding it to develop very small whiskers (Figure 2d) in some of the imaged regions. Recall however that samples were kept under radiation for only small fractions of the above mentioned 15 and 30 days.

All images were used to count whiskers and measure their lengths. While initially we processed images from three different regions, based on their proximity to the source and the average dose level received (100\%, 40\%, or 20\% of the total dose), we found no statistically significant difference in the whisker density or whisker lengths among these regions. Therefore we grouped all data from the irradiated sample together, resulting in 40x3=120 images per irradiation condition, and distinguishing only between the overall total doses received. We address this observation below.

\begin{figure}[ht]
\includegraphics[width=0.4\textwidth]{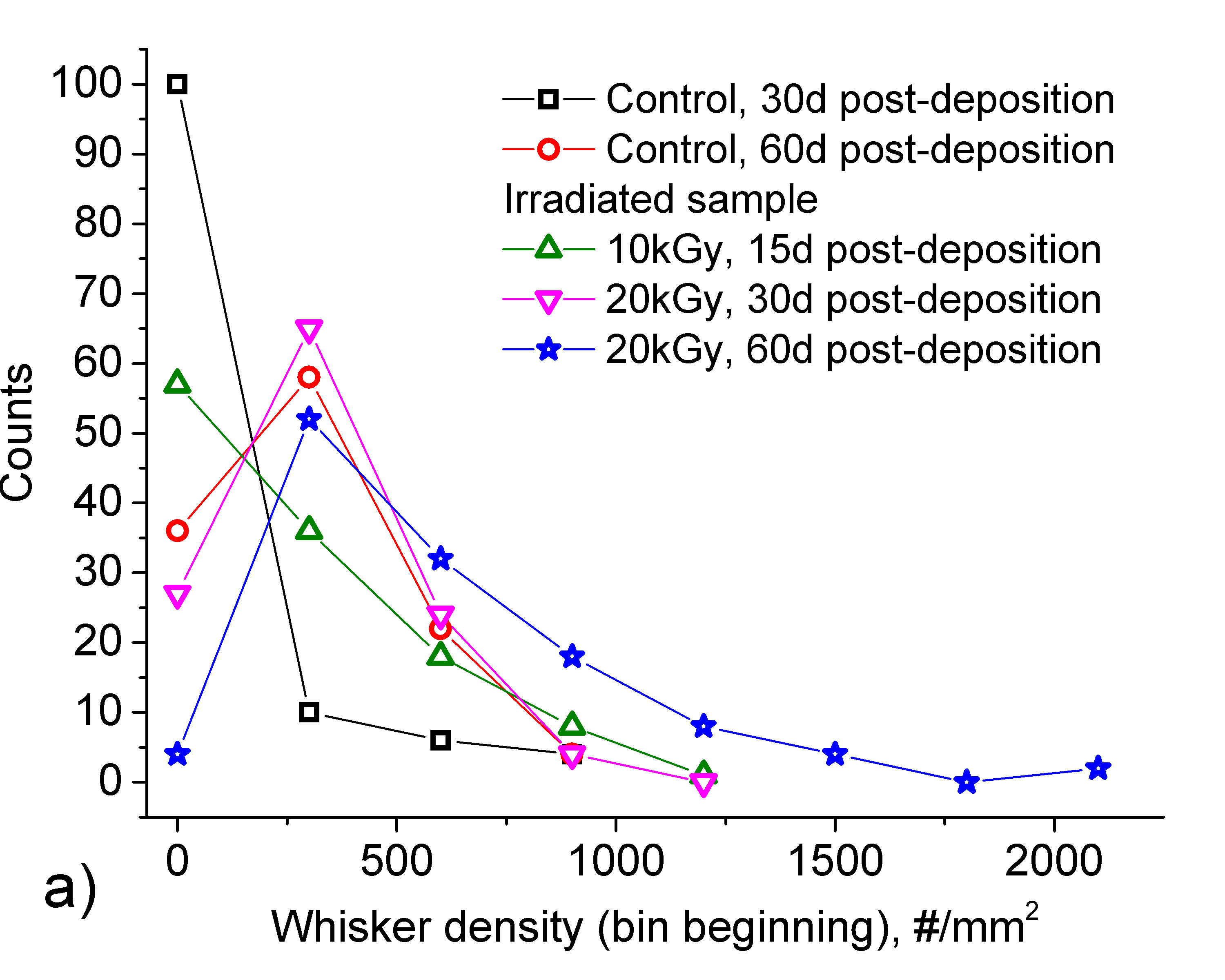}
\includegraphics[width=0.4\textwidth]{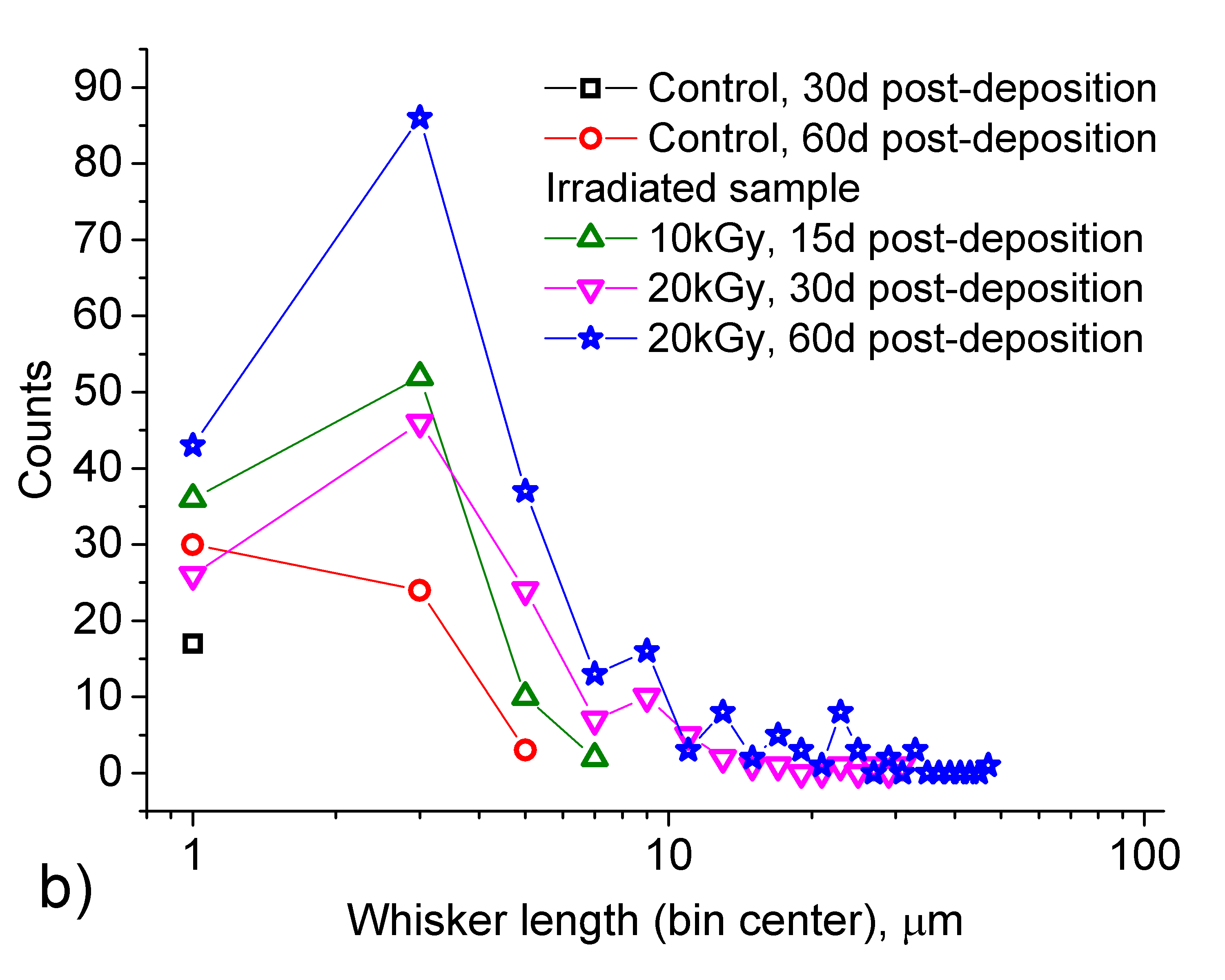}
\caption{Summary whisker statistics for control sample ~30 and 60 days of shelf life (post-deposition), and irradiated sample after 10 and 20kGy doses received, and 30 additional days of shelf life: a) whisker density, 1/mm$^2$; b) whisker lengths, $\mu$m (note log scale of x-axis).\label{fig:stats}}.
\end{figure}

A whisker statistics summary is presented in Fig. \ref{fig:stats} in a form of "frequency counts". Here the total number of counts in Fig. \ref{fig:stats}(a) is 120, corresponding to the number of SEM images processed per irradiation condition (the first point reflects the number of images with no identifiable whiskers). In Fig. \ref{fig:stats}(b) the total number of counts is defined by the total number of whiskers measured, leading to significantly higher count values for the irradiated sample, which developed substantial whiskering after 10kGy, greatly enhanced after the second round of irradiations to 20 kGy.

 \begin{figure}[ht]
\includegraphics[width=0.4\textwidth]{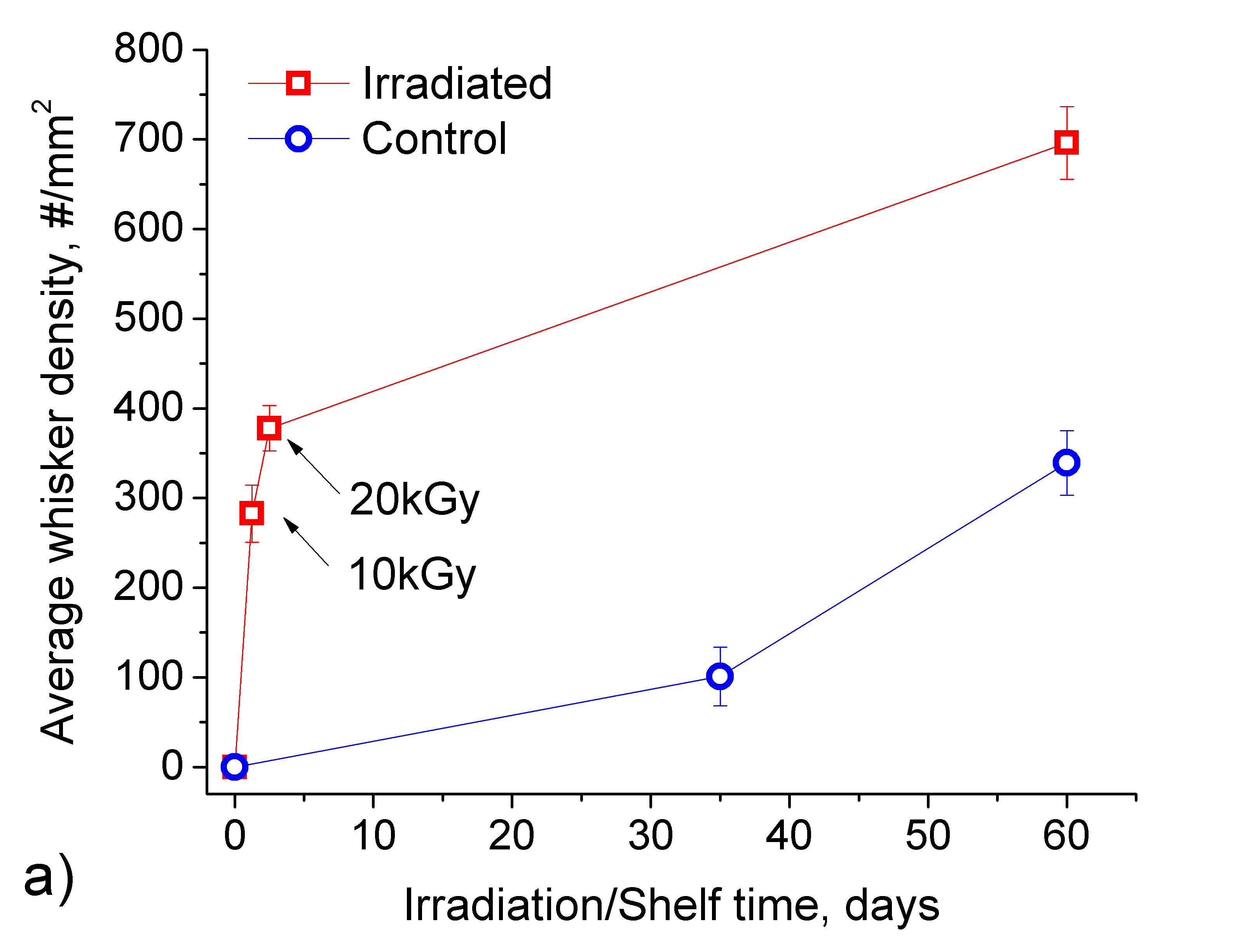}
\includegraphics[width=0.4\textwidth]{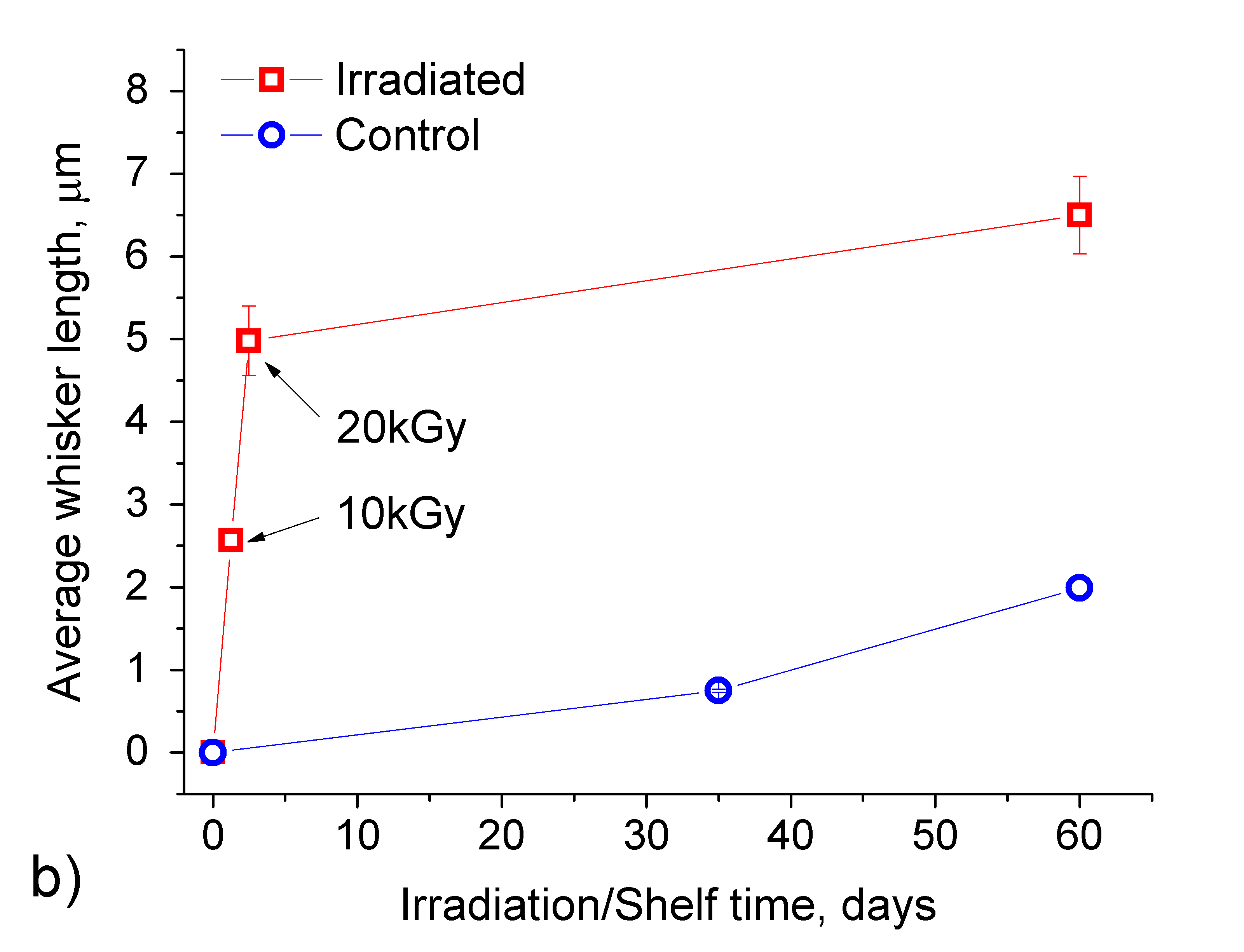}
\caption{Temporal evolution of average a) whisker densities and b) whisker lengths for irradiated and control samples. Note that the radiation treatment times were approximately a factor of ten shorter than the post-deposition times.\label{fig:time}}.
\end{figure}

\begin{table*}[ht]
\centering
%{\small
%\hfill{}
\caption{Summary of average parameters, whisker densities and whisker lengths, for irradiated and control samples.}\label{tab:tab}

\begin{tabular}{|c|c| c| c|c| c| c}
\hline
\multicolumn{4}{|c|}{Irradiated sample} &\multicolumn{2}{|c|}{Control sample}\\
%textbf{Irradiated sample} & \multicolumn{2}{|c|}{\textbf{Control sample}}\\
\hline

Time, days & Dose, kGy & Whisker density, mm$^{-2}$
 & Whisker length, $\mu$m &  Whisker density, mm$^{-2}$ & Whisker length,$\mu$m  \\
 \hline
15      & 10    & 282 $\pm$ 32 &  2.57 $\pm$ 0.12 & &     \\
30      & 20    & 378 $\pm$ 25 &  4.98 $\pm$ 0.42 &101 $\pm$ 33& 0.75 $\pm$ 0.02     \\
60      & 20     & 696 $\pm$ 41 & 6.50 $\pm$ 0.47 & 339 $\pm$ 36 &1.99 $\pm$ 0.12      \\

\hline
%\end{center}
\end{tabular}
%\hfill{}
\end{table*}

Temporal evolution of the average whisker densities and whisker lengths is presented in Fig. \ref{fig:time}. The last point on each of "Irradiated" curves represents the sample irradiated to 20 kGy, held on a shelf for additional 30 days. Error bars in this figure represent statistical uncertainties, calculated as corresponding standard deviations over $\sqrt{N}$, where N=120 is the number of SEM images taken for each sample/condition. While the data have large standard deviations, typical of whisker statistics, the number of processed images does reduce the uncertainty of the average value, making it a more relevant parameter to use for error bars. Numerical values for data summary in Fig. \ref{fig:time} are presented in Table 1; these are used to calculate whisker growth acceleration factor in what follows.

{\it Discussion.} X-ray and gamma-ray induced substrate charging have been studied in connection with various applications (x-ray lithography, degradation of optical devices, and degradation of MOS electronic devices), where such charging can be significant. The effect of radiation on the substrate atoms is that of ionization, knocking off the electrons from the substrates. The electric fields induced through resulting substrate charging fall in the range of MV/cm \cite{shaneyfelt,schwank}, corresponding to those predicted as accelerating whisker growth. \cite{karpov} The field extends over the range of at least tenths of microns; at further distances from the film surface it is screened by ions of surrounding air and depends on the ambient pressure and humidity, as suggested by the Paschen law. \cite{lieberman}

Here we utilized the phenomenon of substrate charging under irradiation to verify the effect of electric field on whisker growth. The observed concomitant darkening of the glass substrate under irradiation is due to a well-known phenomenon of color centers formation in glass, \cite{kreidl} confirming radiation induced changes in the substrate.

As evident from Figures \ref{fig:stats} and \ref{fig:time}, whisker growth was significantly accelerated under irradiation, with both whisker average density and length affected, producing long whiskers, with some examples shown in Figs. \ref{fig:images} (c) and (e). While the control sample also grew some whiskers, they were order of magnitude shorter, almost at the level of "nodules", as presented in Figs. \ref{fig:images} (d) and (f).

A comment is in order regarding our observation of whisker density and length not correlating with the proximity to the radiation source, but dependent only on the total irradiation dose received. It is reasonable to assume that non-uniform irradiation results in the correspondingly nonuniform distribution of substrate charges, seemingly contradicting the observed uniform generation of whiskers. Although unanticipated, that contradiction is easily resolved by taking into effect the field screening due to the electrons in the TCO layer below Sn strip regions. The standard electrostatic calculation (to be published elsewhere along with the observations of nonuniform effects in samples without TCO) shows that free charge carriers in TCO rearrange themselves to fully screen the nonuniform field component while not affecting the average (uniform) field perpendicular to the substrate.  We relate the latter to the observed significant acceleration of whisker growth.

To quantify the effect of electric field on whisker growth we use whisker creation rate
\begin{equation}\nonumber
R=\frac{{\rm number\ of \  whiskers}}{{\rm area}\times {\rm time}}=\frac{{\rm whisker \ density}}{{\rm time}}\end{equation}	
and distinguish between $R_{\rm spon}$ - spontaneous creation rate, with no external field applied, and $R_{\rm stim}$ - stimulated by applied external electric field, generated under gamma-ray irradiation. Next, we define the acceleration factor
	$$a\equiv R_{\rm stim}/R_{\rm spon}.$$	
Taking whisker growth time equal to the irradiation time $t_R\sim 60$ hours \cite{vasko1} required to achieve 20kGy total dose for the irradiated sample, shelf time $t_S$=35 days=840 hours for the control sample, and corresponding average whisker densities from Table \ref{tab:tab}, we arrive at acceleration ratio $a=(378/60)/(101/840)\approx 52$. Evaluating the acceleration factor in terms of relative increase in the average whisker length would result in even higher values for $a\sim 100$. Additional post-irradiation whisker growth appears to follow the rate of the shelf sample. From a practical standpoint that acceleration means growing whiskers under gamma-ray source should take one-two weeks rather than years.

Comparing the obtained results with those previously measured for tin and zinc samples under high-energy electron beams, \cite{vasko1,niraula} we point out roughly 4 times lower values of acceleration factor observed under photon irradiation here. We attribute it to the lower dose rate of the gamma-ray source, resulting in longer irradiation times to achieve the same dose and charge density levels. More experiments with higher dose rate sources are being planned.

{\it Conclusions.} The experimentally observed effect of accelerated Sn whisker growth under gamma-ray irradiation presented here appears quite significant: both whisker densities and lengths are greatly enhanced in the irradiated sample. The effect is attributed to generation of electric charges in the insulating glass substrate supporting the Sn thin films. The charges create an electric field perpendicular to the film surface, thus providing conditions conducive to electrostatically driven whisker growth. The observed acceleration factors of about 50 make this approach promising as a non-destructive readily implementable accelerated life testing tool; the possibility of achieving even higher acceleration factors for higher intensity photon sources of radiation is under investigation.

\section*{Acknowledgements}
We are grateful to G. Davy and A.D. Kostic for insightful discussions and critique of the manuscript.

\end{document}